# A unified formulation of dichroic signals using the Borrmann effect and twisted photon beams

Stephen P Collins[1] and Stephen W Lovesey[1, 2]

1. Diamond Light Source Ltd, Oxfordshire OX11 0DE, UK

2. ISIS Facility, STFC, Oxfordshire OX11 0QX, UK

**Dichroic signals derived from the Borrmann effect and a twisted photon beam with topological charge $l = 1$ are formulated with an effective wavevector. The unification applies for non-magnetic and magnetic materials. Electronic degrees of freedom associated with an ion are encapsulated in multipoles previously used to interpret conventional dichroism and Bragg diffraction enhanced by an atomic resonance. A dichroic signal exploiting the Borrmann effect with a linearly polarized beam presents charge-like multipoles that include a hexadecapole. A difference between dichroic signals obtained with a twisted beam carrying spin polarization (circular polarization) and opposite winding numbers presents charge-like atomic multipoles, whereas a twisted beam carrying linear polarization alone presents magnetic (time-odd) multipoles. Charge-like multipoles include a quadrupole, and magnetic multipoles include a dipole and an octupole. We discuss the practicalities and relative merits of spectroscopy exploiting the two remarkably closely-related processes. Signals using beams with topological charges $l \geq 2$ present additional atomic multipoles.**

Dichroic signals (polarization-dependent spectroscopy) present information on non-magnetic and magnetic ions at an atomic level of detail. The majority of applications, in chemistry, life-sciences and physics, exploit dipole-allowed absorption (E1-E1), because it is a strong event and usually good quality signals can be recorded [1, 2]. Intrinsically weaker signals are elevated in value when dipole events are forbidden by symmetry, or the signals are intentionally selected in the preparation of an experiment for their specific content. Such signals include natural circular, magneto-chiral and non-reciprocal linear dichroism [3, 4, 5]. Recently, two processes that promise to enhance and isolate non-dipolar effects have received prominence in the literature: the Borrmann effect [6, 7, 8], and twisted photon beams that carry non-zero orbital angular momentum [9-15]. We show that these processes are very closely related, and the corresponding dichroic signals offer significant potential to advance the science of materials.

The Borrmann effect, or thick crystal Laue diffraction, arises in high-quality crystals with simple chemical structures, where interference between the incident and diffracted beams sets up a standing wave-field perpendicular to the crystal planes with polarization in the plane of the diffracting atoms. The electric field at the atomic planes vanishes, killing off dipole absorption and allowing 'anomalous transmission'. However, while the field intensity vanishes, the field gradient (now perpendicular to the atomic planes) persists and leads to strong quadrupole absorption (E2-E2) that dominates the recorded spectra, in the absence of significant vibration of the diffracting ions. In a number of experimental reports [6, 7, 8], huge quadrupole enhancements at x-ray K- and L-edges are shown, along with a strong dependence on temperature.

An analogous cancellation of field intensity at the centre of the waist of Laguerre-Gaussian (LG) beams carrying non-zero orbital angular momentum has generated a great deal of interest [12, 13, 14]. Orbital angular momentum in a photon beam is tied to the spatial structure of the wave-front, which is shaped as a helix. A twisted beam with topological (or vortex) charge $l$ carries $l\hbar$ orbital angular momentum directed parallel to the beam axis [10]. Notably, orbital angular momentum is distinct from spin angular momentum associated with circularly polarized radiation and can occur in linearly polarized LG modes. A circularly polarized LG beam possesses spin angular momentum and orbital angular momentum and can exhibit features involving spin-orbit coupling. The absence of field intensity at the central singularity (leading to a dark spot in the beam profile) again kills off dipole absorption from perfectly centred ions, while preserving a significant field gradient and allowing quadrupole transitions.

A formulation of dichroism and related processes uses a resonant contribution in the Kramers-Heisenberg dispersion formula, with Dirac's interpretation of it as a scattering length for a two-step process of photon absorption and emission engaging virtual intermediate states [4]. The scattering length has the form of an amplitude F ~ {V'V} atop an energy denominator that would vanish in the vicinity of a resonance were it not for the lifetime of virtual states. An atomic resonance can be labelled by total angular momentum $J_c$ = $l_c \pm 1/2$, because of strong spin-orbit coupling in the core state. Judd and Ofelt simplified matrix elements like {V'V} that arise in optical transition probabilities. Their result proved to be profoundly important for a majority of studies of electronic transitions within the 4f shell of rare earth ions in solids and solutions, including the evaluation of a rare earth-doped material as a potential laser system [16-19]. Sacrificing some information about intermediate states in {V'V}, in the footsteps of Judd and Ofelt, leads to a factorization F ~ {$\mathbf{H}^K \langle \mathbf{T}^K \rangle$}, where $\mathbf{H}^K$ is a function of the photon variables and an atomic multipole $\langle \mathbf{T}^K \rangle$ encapsulates electronic degrees of freedom in the valence state, and, crucially, it depends on $J_c$ while $\mathbf{H}^K$ does not. The ideas have been applied to electric dipole (E1-E1) and electric quadrupole (E2-E2) absorption events, and the information sacrificed in the Judd-Ofelt prescription can be restored without loosing the advantageous factorization [4, 20-22].

Use of angular brackets for the multipole denotes an expectation value, or time-average, of the enclosed tensor operator of rank *K*, i.e., atomic multipoles depend on the electronic ground-state. Subsequent work confirmed the Judd-Ofelt handling of matrix elements and added sum-rules that are now widely used, whereby the dependence of $\langle \mathbf{T}^K \rangle$ on $J_c$ is exploited to extract expectation values of occupation numbers, the spin-orbit interaction, and spin and orbital angular momenta from integrated signals [2, 23, 24]. The discrete symmetries of $\langle \mathbf{T}^K \rangle$ are parity and time-reversal, with multipoles parity-even for (E1-E1) and (E2-E2) absorption events and time-odd (time-even) for *K* odd (even).

The quantity extracted from experiments is the absorption coefficient μ(E) where E = ℏqc is the photon energy. Values of μ(E) and the previously mentioned scattering length, denoted here by $f$, are related by the so-called optical theorem,

$$\mu(E) = (4\pi/q)\, \text{Im}.f.$$

The Kramers-Heisenberg dispersion formula yields,

$$f = -(r_e/m)\, (\{V'V\}/[E - \Delta + i\Gamma/2]),$$

for E in the vicinity of an atomic resonance with an energy $\Delta$ and a lifetime $\propto \hbar/\Gamma$. As its name implies, $f$ has the dimension of length while μ(E) has the dimension of area ($r_e \approx 0.282$ $10^{-12}$ cm, $mc^2 \approx 511$ keV). Instrumentation and experimental methods for conventional dichroism are reviewed by van der Laan and Figueroa [2].

The (E2-E2) photon tensor $\mathbf{H}^K$ depends on the polarization vector of the primary beam $\boldsymbol{\epsilon}$ ($\boldsymbol{\epsilon}'$ secondary beam) and wavevector $\mathbf{q}$ ($\mathbf{q}'$), and the topological charge and winding number in the case of a twisted beam. Circular polarization in the primary beam - its spin angular momentum - imposes selection rules on $\mathbf{H}^K$ that can exclude some atomic multipoles from a dichroic signal. Likewise, we demonstrate that the winding number in a twisted beam imposes potentially useful selection rules that identify the physical properties of multipoles that can be observed. The potential usefulness of dichroic signals created by a twisted beam has been illustrated with numerical simulations of electronic spectra for cuprates, manganites and ruthenates [25]. Our new findings flow from intuitive reasoning, backed by explicit calculations, that dichroism using a twisted beam and dichroism created by the Borrmann effect can be mapped to an (E2-E2) event; both forms of dichroism are described by existing theory F ~ $\{\mathbf{H}^K \langle \mathbf{T}^K \rangle\}$ with effective wavevectors replacing true wavevectors in $\mathbf{H}^K$. In consequence, established sum rules for (E2-E2) events are preserved in the novel dichroic signals under discussion. The effective wavevector is complex in the case of a twisted beam, and purely real for the Borrmann effect. In the latter case, time-even (charge-like) multipoles with rank up to $K = 4$ (hexadecapole) can be observed with linear polarization. In an application of a twisted beam, the difference $\Delta F$ in dichroic signals observed with opposite signs for the winding number appears to be potentially useful, because subtraction of the two signals selects multipoles with specific properties. The difference signal $\Delta F$ contains magnetic multipoles ($K = 1$ & 3) for linear polarization and non-magnetic multipoles ($K = 0$ & 2) for circular polarization.

**Results**

The theory of absorption from a single plane-wave is well-established. We show that non-plane-wave spectroscopies of interest, using the Borrmann effect and twisted beams, can be described by simple extensions of existing knowledge of dichroic signals. We begin with a rather simple approach, but illustrate later a formal relation to normal dichroic signals that generates valuable selection rules for the new spectroscopies.

Consider first (E1-E1) scattering, which produces the leading-order term in normal plane-wave absorption, and enters both the Borrmann and twisted beam cases when absorbing ions are displaced from ideal, central positions. Factorization of photon and electronic variables that has been mentioned appears as a contraction of two spherical tensors in the scattering amplitude,

$$F(E1) = \sum_{K,Q} (-1)^Q X^K_{-Q} \Psi^K_Q,$$

with,

$$X^K_Q = \sum_{\alpha,\beta} \epsilon'_\alpha \epsilon_\beta (1\alpha\, 1\beta | KQ),$$

in which $(1\alpha\, 1\beta | KQ)$ is a standard Clebsch-Gordan coefficient [22]. The electronic structure factor $\Psi^K_Q$ in F(E1) is a suitable linear combination of atomic multipoles of rank $K$. The photon tensor $X^K_Q$ depends exclusively on photon polarization (electric field) vectors, $\boldsymbol{\epsilon}$ and $\boldsymbol{\epsilon}'$, and so absorption by a coherent superposition of beams is given by replacing polarization vectors with the sum of component fields, which vanishes in the ideal extremes of spectroscopy using the Borrmann effect and twisted beams.

Quadrupole absorption is more interesting as it is driven by the field gradient, which depends on polarizations and wavevectors. The corresponding photon tensor couples two second-rank tensors to form a resultant up to rank-four, coupling to atomic multipoles in $\Psi^K_Q$ up to the same rank. One finds [4, 21],

$$H^K_Q = \sum_{r,s} h(r)\, h'(s)\, (2r\, 2s | KQ), \qquad (1)$$

$$h(r) = \sum_{\alpha,\beta} \epsilon_\alpha \hat{q}_\beta (1\alpha\, 1\beta | 2r) \text{ and } h'(s) = \sum_{\alpha,\beta} \epsilon'_\alpha \hat{q}'_\beta (1\alpha\, 1\beta | 2s).$$

Anticipating that wavevectors depicted in Fig. 1a are soon replaced by effective wavevectors, to be determined, wavevectors in h($r$) and h'($s$) are labelled by $\hat{\mathbf{q}}$ and $\hat{\mathbf{q}}'$ for clarity at the moment. Since h($r$) relates to one of the matrix elements, V, in the two-step process of photon absorption and emission, it is this object that has to be summed over, with appropriate weights, $\alpha_j$, to account for multiple wave components. A generalized value is taken to be,

$$h(r) \to \sum_j \alpha_j\, h_j(r).$$

We conclude that an existing framework for single plane-wave scattering can be generalized to account for absorption by multiple plane-wave components, each with distinct polarization. However, a significant simplification occurs when all components share the same polarization. In that case, the sum over j in the generalized h($r$) is identical to the single-wave case if the wavevector is replaced with an effective wave vector, obtained by summing over all contributing waves, namely,

$$\hat{\mathbf{q}} \to \boldsymbol{\kappa} = \sum_j \alpha_j\, \mathbf{q}_j, \text{ and choose } \boldsymbol{\kappa} \cdot \boldsymbol{\kappa}^* = 1.$$

One can describe all of the quadrupole processes in the present work using this substitution.

In the Borrmann case, depicted in Fig. 1b, summation is over two waves with opposite signs giving $\kappa_b \propto (\mathbf{q} - \mathbf{q}') \propto \sin\theta\,(0, 1, 0)$. A twisted beam with topological charge $l = 1$ is very similar, but rather than comprising two components there is a continuum of wavevectors around the cone depicted in Fig. 1c. Moreover, the phase varies linearly with azimuth, $\varphi$, depending on the winding number $n$, as described in detail later. The resulting effective wavevector is obtained by integrating around the cone, in the paraxial limit ($\theta \to 0$), and $\kappa_t \propto \sin\theta\,(-i, 1, 0)$ for $n = +1$, and $\kappa_t \propto \sin\theta\,(i, 1, 0)$ for $n = -1$.

The relationship between the Borrmann and twisted beam cases is now very clear. Both have effective wavevectors that are perpendicular to the average beam direction $(0, 0, 1)$ in Fig. 1a, and both scale with $\sin\theta$. However, $\kappa_b$ is real, while a complex $\kappa_t$ is reminiscent of a complex polarization vector often used to represent circular polarization, with spin angular momentum $\pm 1$, consistent with an interpretation of a twisted beam in terms of orbital angular momentum. Finally, we note that the Borrmann case is identical to a linear combination of twisted beams with $n = \pm 1$. Armed with these effective wavevectors, $\kappa_b$ and $\kappa_t$, we can compute photon tensors for all cases of interest, and therefore determine which sample properties can be obtained by the corresponding measurement. Results in Table 1 for $H^K_Q$ use $\hat{\mathbf{q}} = \kappa$ and $\hat{\mathbf{q}}' = -\kappa^*$. An expanded tabulation, including results for $X^K_Q$, are included in the Supplementary Information.

**Borrmann effect.** The standard electric multipole expansion of the electron-photon interaction V treats the product of the electron position ($\mathbf{r}$) and photon wavevector ($\mathbf{q}$) as a small quantity, i.e., ($\mathbf{r} \cdot \mathbf{q}$) << 1 [4]. The Borrmann effect is similar with two adaptions: (i) the electric field for a single travelling wave is replaced by that of two waves with one along the incident beam direction, $\mathbf{q}$, and the other, of equal amplitude, along the diffracted beam direction, $\mathbf{q}'$, where the two waves are phased so as to give zero-field at the diffracting planes and (ii) absorption is by ions displaced from their ideal positions by a small distance $\mathbf{u}$. The resulting form of V for the Borrmann effect is then,

$$V \propto \mathbf{r} \cdot \boldsymbol{\epsilon}\,[\exp(i\{\mathbf{r}+\mathbf{u}\}\cdot \mathbf{q}) - \exp(i\{\mathbf{r}+\mathbf{u}\}\cdot \mathbf{q}')] = i\mathbf{r}\cdot\boldsymbol{\epsilon}\,(\kappa\cdot\{\mathbf{r}+\mathbf{u}\}) + \ldots,$$

where $\kappa \propto \mathbf{q} - \mathbf{q}'$. Here, the leading (dipole) term in the normal absorption case has vanished, leaving two terms of the same order. The first of these is a quadrupole term $\mathbf{r}\,(\kappa \cdot \mathbf{r})$, and the second is dipolar absorption that arises due to the atomic displacement [6, 7, 8].

To complete the implied mapping of the Borrmann effect to the standard (E2-E2) scattering amplitude we use Cartesian coordinates in Fig. 1a, with our visualization of wavevectors for the effect depicted in Fig. 1b. Cartesian forms of the wavevectors in Fig. 1b are $\mathbf{q} = q(0, -\sin\theta, \cos\theta)$ and $\mathbf{q}' = q(0, \sin\theta, \cos\theta)$ which leaves $\mathbf{q} - \mathbf{q}'$ parallel to the y-axis. We choose an effective wavevector $\kappa_b = (0, 1, 0)$ with $\boldsymbol{\epsilon} = (1, 0, 0)$ to evaluate the photon tensor $H^K_Q$. The amplitude in the corresponding (E2-E2) scattering length is,

$$F(E2) = \sin^2\theta \sum_{K,Q} (-1)^{K+Q} H^K_{-Q}\,\Psi^K_Q, \tag{2}$$

with an electronic structure factor,

$$\Psi^K{}_Q = \sum_{\mathbf{d}} \langle T^K{}_Q \rangle_{\mathbf{d}}, \qquad (3)$$

and the sum is over sites **d** in a unit-cell used by resonant ions. The Discussion includes specific examples of electronic structure factors for magnetic materials.

**Twisted beam.** The case of twisted beam is almost identical to that of the Borrmann effect, except that the effective wavevector is now determined by integrating over a continuum of states, each lying on a cone depicted in Fig. 1c. A phase is determined by the azimuthal angle φ around the cone. For a topological charge $l$,

$$\boldsymbol{\kappa} \propto \sin\theta \int_0^{2\pi} d\varphi \exp(il\varphi) (-\sin\varphi, -\cos\varphi, 0).$$

Taking $l = 1$ we choose effective wavevectors $\boldsymbol{\kappa}_t = (-i, 1, 0)/\sqrt{2}$ for $n = +1$, and $\boldsymbol{\kappa}_t = (i, 1, 0)/\sqrt{2}$ for $n = -1$.

**Theory**. It was previously shown that an alternative version of the photon tensor (1) is helpful in exposing selection rules in normal dichroic signals and resonance enhanced Bragg diffraction [4], and so it is in the present discussion of signals derived from the Borrmann effect and twisted beams. A derivation of results,

$$H^K{}_Q = 5(-1)^K \sum_{K'K''} \{\Pi^{K'} \otimes X^{K''}\}^K{}_Q [(2K'+1)(2K''+1)]^{1/2} \begin{Bmatrix} K' & K'' & K \\ 1 & 1 & 2 \\ 1 & 1 & 2 \end{Bmatrix}, \qquad (4)$$

with coupled tensors,

$$\Pi^{K'}{}_{Q'} = \{\hat{\mathbf{q}}' \otimes \hat{\mathbf{q}}\}^{K'}{}_{Q'}, \quad X^{K''}{}_{Q''} = \{\boldsymbol{\epsilon}' \otimes \boldsymbol{\epsilon}\}^{K''}{}_{Q''},$$

$$\{\Pi^{K'} \otimes X^{K''}\}^K{}_Q = \sum_{Q',Q''} \Pi^{K'}{}_{Q'} X^{K''}{}_{Q''} (K'Q' \, K''Q'' | KQ).$$

for a twisted beam is provided in the Supplementary Information. Necessary recoupling of angular momenta is thoroughly reviewed by Balcar and Lovesey [22]. Tensor ranks $K$, $K'$, $K''$ are subject to triangular conditions on arguments in rows and columns of the 9-j symbol, and the symbol vanishes unless $(K + K' + K'')$ is an even integer. The specific form of $\Pi^{K'}{}_{Q'}$ for a twisted beam can be deduced from the effective wavevector generated from Fig. 1c for $l = 1$ using the definition,

$$\Pi^{K'}{}_{Q'} = \sum_{\alpha,\beta} \hat{q}'_\alpha \hat{q}_\beta (1\alpha \, 1\beta | K'Q').$$

The desired results are obtained from $\hat{\mathbf{q}} = \boldsymbol{\kappa}_t$ and $\hat{\mathbf{q}}' = -(\hat{\mathbf{q}})^*$. Specifically, spherical components $\hat{q}_{+1} = 0$, $\hat{q}_{-1} = -i$, $\hat{q}_0 = 0$ and $\hat{q}'_\alpha = -\hat{q}_{-\alpha}$ using $\boldsymbol{\kappa}_t = (-i, 1, 0)/\sqrt{2}$ for $n = +1$, while $\hat{q}_{+1} = -i$, $\hat{q}_{-1} = 0$, $\hat{q}_0 = 0$ and $\hat{q}'_\alpha = -\hat{q}_{-\alpha}$ using $\boldsymbol{\kappa}_t = (i, 1, 0)/\sqrt{2}$ for $n = -1$. Thus, Clebsch-Gordan coefficients in $\Pi^{K'}{}_{Q'}$ are of the form $(11 \, 1-1|K'Q')$ and $(1-1 \, 11|K'Q')$, and vanish unless $Q' = 0$. Primary and secondary polarization vectors $\boldsymbol{\epsilon}$ and $\boldsymbol{\epsilon}'$ are not orthogonal to the complex wavevector $\boldsymbol{\kappa}_t$.

Valuable selection rules flow from the result $\Pi^{K'}_{Q'} = (lnl{-}n | K'Q')$ deduced for topological charge $l = 1$ and winding number $n = \pm l$. One has $(lnl{-}n | K'Q') \propto \delta_{Q',0}$, and $(ln\,l{-}n | K'\,0) = (-1)^{K'}(l{-}n\,ln | K'\,0)$. A first selection rule is $Q = Q''$ in (4), with allowed projections of atomic multipoles $\langle T^K_Q \rangle$ in the electronic amplitude F(E2), defined in (2), actually selected by the polarization factor $X^{K''}_{Q''}$. Polarization vectors lie in the x-y plane of Fig 1a, to a good approximation. Circular polarization picks out $K'' = 1$ and $Q'' = 0$ in $X^1_{Q''}$, while linear polarization picks out $K''$ even. Signal selection, using the change in sign of F(E2) with winding number $n = \pm 1$, and circular polarization gives access to charge-like atomic multipoles, while magnetic multipoles are accessed with linear polarization. Since $\Pi^{K'}_0$ is unchanged by a change in sign of $n$ for $K'$ even and reverses its sign with respect to the sign of $n$ for $K'$ odd, selection of components of F(E2) on the basis of the winding number, a dichroic signal labelled $\Delta$F, uses $K' = 1$ and $(K + K'')$ odd. The selection rules are thus, $K$ even for circular polarization and $K$ odd for linear polarization for the multipole ranks in the atomic multipole $\langle T^K_Q \rangle$.

Atomic multipoles in $\Delta F = \{F(E2; n = +1) - F(E2; n = -1)\}$ observed with circular polarization are time-even multipoles $\langle T^0_0 \rangle$ & $\langle T^2_0 \rangle$ that are purely real. The result follows from the triangular condition on $K, K', K''$ in the 9j-symbol with $K' = K'' = 1$. Corresponding photon tensors are summarized in Table 1, and Table 4 is a complete listing of $H^K_Q$. Linear polarization presents magnetic multipoles $\langle T^1_0 \rangle$ & $\langle T^3_Q \rangle$ with projections $Q = 0$ & $\pm 2$ in the difference signal $\Delta F$, and photon tensors therein are found in Table 1.

For the Borrmann effect $\hat{\mathbf{q}} = (0, 1, 0) = \boldsymbol{\kappa}_b$ with $\hat{q}_{+1} = \hat{q}_{-1} = -i/\sqrt{2}$, and $\hat{q}'_\alpha = -\hat{q}_{-\alpha}$. In consequence,

$$\Pi^{K'}_{Q'} = (1/2)\{\delta_{K',2}[\delta_{Q',+2} + \delta_{Q',-2} + \sqrt{(2/3)}\,\delta_{Q',0}] + (2/\sqrt{3})\,\delta_{K',0}\,\delta_{Q',0}\}, \quad (5)$$

which is used to calculate photon tensors gathered in Table 1. With $K'$ even and linear polarization ($K'' = 0, 2$) the Borrmann effect engages charge-like electronic multipoles ($K$ even). The contribution $K' = 2$ presents the hexadecapole $\langle T^4_Q \rangle$ with projections $Q = 0, \pm 2, \pm 4$, but $H^4_{\pm 2} \propto [1 - \exp(2i\phi)] = 0$ for linear polarization parallel to the x-axis in Fig. 1a.

**Discussion**

Our principal result is that the expression (2) for the amplitude of an (E2-E2) dichroic signal is valid for signals created by the Borrmann effect and a twisted beam. The corresponding amplitudes are obtained from a mapping that uses an effective wavevector, $\boldsymbol{\kappa}$, depicted in Figs. 1b & 1c. Mapping of the Borrmann effect is achieved with real $\boldsymbol{\kappa}_b = (0, 1, 0)$, whereas a mapping of the twisted beam requires complex $\boldsymbol{\kappa}_t = (\pm i, 1, 0)/\sqrt{2}$ and signs to denote the winding number of the primary beam. Thereafter, it is straightforward to delineate selection rules for the new dichroic signals, by appealing to prior knowledge about the properties of parity-even atomic multipoles $\langle \mathbf{T}^K \rangle$ used to describe the electronic state of ions engaged in conventional dichroic signals, and, also, the Bragg diffraction of x-rays enhanced by an atomic resonance [4]. Atomic multipoles of even rank are charge-like and those with odd rank are magnetic. The Borrmann effect presents charge-like multipoles with rank $K \leq 4$, while a twisted beam presents either charge-like (circular polarization) or magnetic multipoles (linear polarization) depending on polarization in the primary beam. Previous numerical simulations of dichroic signals using a twisted beam, accomplished for various materials, underscore the potential value of the technique [25].

--------------------------------------------------------------------------------

**Table 1**. **Photon Tensor $H^K$**. The tensor $H^K_Q$ is derived from either (1) or (4). F(E2) is defined (2) and it contains $\sin^2\theta$ with $\theta$ defined in Fig. 1a: Borrmann effect $\hat{\mathbf{q}} = \boldsymbol{\kappa}_b = (0, 1, 0)$ and $\hat{\mathbf{q}}' = -\boldsymbol{\kappa}_b$ with $\boldsymbol{\epsilon} = (1, 0, 0)$: Twisted beam, winding number $n = (\pm 1)$, with $\hat{\mathbf{q}} = \boldsymbol{\kappa}_t$ and $\hat{\mathbf{q}}' = -\boldsymbol{\kappa}_t{}^*$ using $\boldsymbol{\kappa}_t = (-i, 1, 0)/\sqrt{2}$ for $n = +1$, and $\boldsymbol{\kappa}_t = (i, 1, 0)/\sqrt{2}$ for $n = -1$. A difference signal $\Delta F = \{F(E2; n = +1) - F(E2; n = -1)\}$ contains $H^K_Q$ in the table. In the case of a twisted beam with circular polarization, defined by the Stokes parameter $P_2$, a difference signal $\{\Delta F(P_2 = +1) - \Delta F(P_2 = -1)\} = (4 \sin^2\theta \, H^K_0 \langle T^K_0 \rangle)$, using the tabulated values for $K = 0$ & 2. See also entries in Table 4 in SI.

*Borrmann effect*: linear polarization

$H^0_0 = -(1/2)(1/\sqrt{5})$, $H^2_0 = -(1/\sqrt{14})$, $H^4_0 = -(1/2)(1/\sqrt{70})$, $H^4_{\pm 4} = (1/4)$

*Twisted beam*: linear polarization subtends an angle $\phi$ with the x-axis

$H^1_0 = (1/\sqrt{10})(\pm 1)$, $H^3_0 = (1/2)(1/\sqrt{10})(\pm 1)$, $H^3_{\pm 2} = -\{(1/4)(1/\sqrt{3}) \exp(\pm 2i\phi)\}(\pm 1)$

*Twisted beam*: circular polarization

$H^0_0 = -(1/12)\sqrt{5}$, $H^2_0 = -(1/6)\sqrt{(7/2)}$

--------------------------------------------------------------------------------

Sum-rules for integrated signals follow from known properties of $\langle \mathbf{T}^K \rangle$ [2, 4]. Absorption at a K-edge will be useful for 3d and 5d transition ions. Atomic multipoles are functions of electronic orbital degrees of freedom at the K-edge, with $\langle \mathbf{T}^1 \rangle$ proportional to orbital angular momentum $\langle \mathbf{L} \rangle$ [27]. Spin and orbital degrees of freedom in ions contribute at L-edges, useful for rare earth and actinide compounds. Spin contributions to $\langle \mathbf{T}^K \rangle$ are exposed by taking a difference of integrated dichroic signals at two L-edges, say [4].

By way of an example, consider the osmate $NaOsO_3$ that forms an orthorhombic lattice with four formula units per cell, and superstructure as in the $GdFeO_3$-type perovskite [28]. Cooling through $T_c \approx 410$ K triggers a continuous, second-order phase-transition to an antiferromagnetic insulator with no change to the lattice. The electronic structure factor (3) is,

$$\Psi^K_Q(NaOsO_3) = [1 + \sigma_\theta (-1)^Q] [\langle T^K_Q \rangle + \sigma_\theta (-1)^K \langle T^K_{-Q} \rangle],$$

for Os ions using sites 4a in the magnetic space-group Pn′ma′. Sites 4a possess inversion symmetry and no more. The label $\sigma_\theta = (-1)^K$ is the time signature of $\langle T^K_Q \rangle$. By taking $\sigma_\theta = -1$ a ferromagnetic motif of dipoles ($K = 1$) parallel to the b-axis is allowed. However, magnetic signals require $Q$ odd, and these projections are forbidden in dichroism created with a twisted beam and linear polarization. Charge-like signals ($K$ even) are allowed for both types of dichroism under discussion.

Neptunium dioxide undergoes an uncommon form of electronic phase transition at about 25 K [29], which could be further investigated by measuring dichroism at Np $L_2$ and $L_3$ absorption edges. Symmetry of the fluorite structure $Fm\bar{3}m$ (#225) is reduced to $Pn\bar{3}m$ (#224) at the transition, and there is no experimental evidence in favour of a long-range order of Np magnetic dipoles. The electronic structure factor is,

$$\Psi^K{}_Q(NpO_2) = [1 + (-1)^Q] \,[\langle T^K{}_Q \rangle + (-1)^K \langle T^K{}_{-Q} \rangle],$$

and restrictions imposed by site symmetry are sever, even though it includes inversion and allowed Np multipoles parity-even. Magnetic dipoles $\langle \mathbf{T}^1 \rangle$ are forbidden, as is the quadrupole $\langle T^2{}_0 \rangle$. In addition, $\langle T^K{}_{-Q} \rangle = \langle T^K{}_Q \rangle^* = (-1)^{K+p} \langle T^K{}_Q \rangle$ with $Q = 2p$ and $p$ an integer. In consequence, $\Psi^K{}_Q(NpO_2)$ can be different from zero for $p$ even, and with the Borrmann effect it is possible to measure a signal due to hexadecapoles $\langle T^4{}_0 \rangle$ and $\mathrm{Re}.\langle T^4{}_{\pm 4} \rangle$.

Our unified picture of spectroscopy using the Borrmann effect and a twisted beam provides considerable physical insight into both processes, and highlights their common features and differences. A twisted beam is more versatile in terms of the accessible multiples. However, the fact that several experimental spectra have been published for the Borrmann case [6, 7, 8], while x-ray spectroscopy using a twisted beam is yet to take hold, suggests that the latter is intrinsically difficult. While producing x-ray twisted beams is non-trivial [15, 31], it is possible that a more fundamental difficulty arises from the need to have large cone angles (large θ) while maintaining a wide enough beam waist for normal dipole absorption to be minimized. In the Borrmann case, the angles are very large, leading to large effective wavevectors (large field gradients), and the volume of vanishing field intensity is then so small that even atomic thermal fluctuations dominate measured spectra. A ratio of the (time-averaged) quadrupole to dipole (E2/E1) signals yields an enhancement factor for the Borrmann effect $\propto \lambda^2/\langle u^2 \rangle$, where $\lambda$ is the photon wavelength and u is an atomic displacement [6, 7, 8]. Similar sub-Angstrom positioning would be required with a twisted beam, even if such large angles could be realized. Smaller cone angles increase the waist size but reduce the field gradient. Although the dipole and quadrupole absorption scale together in the ideal case, for small angles even a minute background contamination might swamp the signal. Experimental verification of x-ray spectroscopy using a twisted beam therefore represents an important and interesting challenge, which is perhaps made easier by considering longer-wavelength soft x-rays.

In summary.

- All contributions to dichroic signals that are symmetric with respect to rotation about the z-axis in Fig. 1a (such as normal absorption with circular light) have vanishing tensor components for all non-zero projections.
- Odd-rank atomic multipoles are time-odd (magnetic) multipoles and *vice versa*.

- The Borrmann effect with linear polarization probes only time-even (charge-like) multipoles.
- A twisted beam with linear polarization gives no dipole contribution.
- A twisted beam probes both even (charge) and odd (magnetic) multipoles.
- For linear polarization, reversing the winding number *n* reverses the magnetic contribution.
- Amplitudes F(E2) in equation (2) for the Borrmann and twisted beams cases have pre-factors that scale with $\sin^2\theta$ and therefore become vanishingly small for small cone-angles.

**Methods**.

Radiation is treated classically in the paraxial approximation. The spatial spread of electronic wavefunctions is assumed to be small compared to the waist of the primary beam. An LG polarization vector lies in the plane normal to the beam's direction of propagation, to a good approximation. The electric field of a LG beam has a small vector component along the propagation axis and it has been shown to be safely neglected [30].

Electronic degrees of freedom, charge, spin and orbital angular momenta, in the ground-state of ions engaged in dichroic signals are treated by spherical tensor operators with defined discrete symmetries [4]. Balcar and Lovesey [22] review recoupling of angular momenta required to achieve the result (4) for the product of matrix elements in the amplitude.

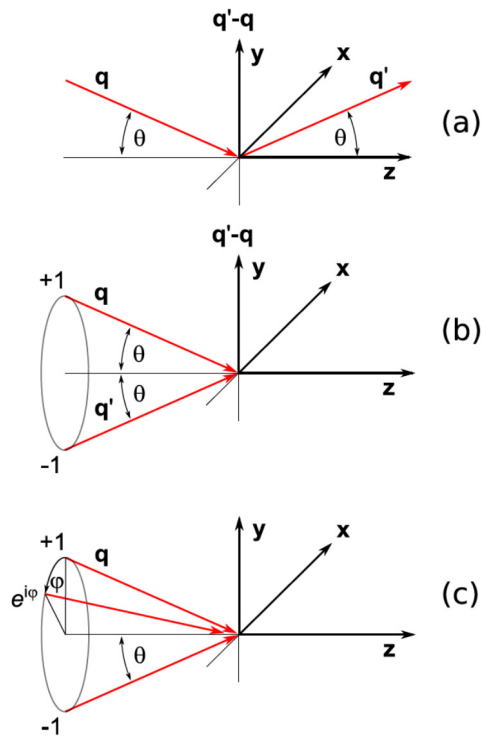

**Figure 1**. **Coordinate systems**. (a, b): Cartesian coordinates (x, y, z) using the x-axis normal to the plane of scattering that contains wavevectors **q** (primary) and **q'** (secondary) with **q** • **q'** = $q^2 \cos(2\theta)$ (c): a twisted beam ($l = 1$) can be viewed as a generalization of the Borrmann case, with a continuum of beams and a phase that varies continuously around a cone.

**Acknowledgements** One of us (SWL) is grateful to Dr Valerio Scagnoli for an early introduction to twisted photon beams.

**Author Contributions** Authors (SPC) and (SWL) have contributed equally to the work.

**Additional Information**

**Supplementary information** accompanies this paper at

**Competing financial interests**: The authors declare no competing financial interests.

**Supplementary information**

**A unified formulation of dichroic signals using the Borrmann effect and twisted photon beams**

Stephen P Collins[1] and Stephen W Lovesey[1, 2]

1. Diamond Light Source Ltd, Oxfordshire OX11 0DE, UK

2. ISIS Facility, STFC, Oxfordshire OX11 0QX, UK

A scattering length $f$ derived from the Kramers-Heisenberg dispersion function is,

$$f = -F/(E - \Delta + i\Gamma/2), \quad (A1)$$

where E is the photon energy, and $\Delta$ and $\Gamma$ are the energy and total width of the atomic resonance labelled $\eta$. The amplitude F = {V'$_{2\eta}$ V$_{1\eta}$} where matrix elements of the electron-photon interaction operator V between states 1 & 2 and the intermediate state V'$_{2\eta}$ & V$_{1\eta}$ account for photon creation and annihilation [32]. V is proportional to the photon polarization vector $\boldsymbol{\epsilon}$. In consequence, F contains a product $\epsilon'_\alpha \epsilon_\beta$ that is usefully expressed through a tensor product,

$$X^{K''}{}_{Q''} = \sum_{\alpha,\beta} \epsilon'_\alpha \epsilon_\beta \, (1\alpha \, 1\beta | K''Q'') = \{\boldsymbol{\epsilon}' \otimes \boldsymbol{\epsilon}\}^{K''}{}_{Q''}. \quad (A2)$$

The Clebsch-Gordan coefficient and Wigner 3-j symbol in (A2) are standard with,

$$(a\alpha b\beta | KQ) = (-1)^{-a+b-Q} \sqrt{(2K+1)} \begin{pmatrix} a & b & K \\ \alpha & \beta & -Q \end{pmatrix}.$$

One finds $X^{K''}{}_{-Q''} = (-1)^{K''+Q''} (X^{K''}{}_{Q''})^*$ and,

$$X^0{}_0 = -(1/\sqrt{3})\,[\boldsymbol{\epsilon}' \cdot \boldsymbol{\epsilon}], \; \mathbf{X}^1 = (i/\sqrt{2})\,[\boldsymbol{\epsilon}' \times \boldsymbol{\epsilon}], \; X^2{}_0 = (1/\sqrt{6})\,[3\epsilon'_0\epsilon_0 - \boldsymbol{\epsilon}' \cdot \boldsymbol{\epsilon}],$$

$$X^2{}_{+1} = (1/\sqrt{2})\,[\epsilon'_0\epsilon_{+1} + \epsilon'_{+1}\epsilon_0], \; X^2{}_{+2} = \epsilon'_{+1}\epsilon_{+1}. \quad (A3)$$

Our Cartesian coordinate scheme is depicted in Fig. 1a. For a normal dichroic signal and linear polarization the required values of $\mathbf{X}^{K''}$ use $\boldsymbol{\epsilon}' \cdot \boldsymbol{\epsilon} = 1$, with $X^2{}_0 = -(1/\sqrt{6})$, $X^2{}_{+1} = 0$ and $X^2{}_{\pm 2} = 1/2$. Upon averaging over circular polarization in the primary photon beam $(\mathbf{X}^1)_{av} = -(1/\sqrt{2})\,\hat{\mathbf{q}}\,P_2$, where $P_2$ is a Stokes parameter for circular polarization [4] and $\hat{\mathbf{q}} = (0, 0, 1)$.

We calculate the the value of F produced by the interaction of twisted radiation with ions, and adopt the standard assumptions. A dipole matrix element of the type needed in F has been calculated by Alexandrescu *et al*. with the same assumptions [11]. Radiation is treated classically in the paraxial approximation. The spatial spread of electronic states is assumed to be small compared to the waist $w$ of the twisted beam. In these circumstances the electric field **E** can be expressed in terms of solid spherical-harmonics $\Re^l_n(\mathbf{b})$ with an argument **b** proportional to the transverse component $\mathbf{r}_\perp$ of the position of an electron. The angular orientation of **b** is carried by a spherical harmonic in $\Re^l_n(\mathbf{b})$. For a transverse component $\mathbf{r}_\perp$ the topological charge and its projection must satisfy $l + n$ even and,

$$\mathbf{E} \propto \boldsymbol{\epsilon}\, \Re^l_n(\mathbf{b}), \tag{A4}$$

with $n = \pm l$ and $\mathbf{b} = \mathbf{r}_\perp/w$. The polarization vector $\boldsymbol{\epsilon}$ and $\mathbf{r}_\perp$ are confined to the plane normal to the direction of propagation of the beam, which is taken to be the z-axis in Fig. 1 of the main text. The proportionality factor in (A4) is purely real. The corresponding dipole interaction operators are,

$$V \propto \mathbf{r} \cdot \boldsymbol{\epsilon}\, \Re^l_n(\mathbf{b}), \text{ and } V' \propto V^* \text{ using a polarization vector } \boldsymbol{\epsilon}', \tag{A5}$$

with the electron position $\mathbf{r} \propto \Re^1(\mathbf{r})$ measured relative to an origin at **R**, giving $w\mathbf{b} = \mathbf{R}_\perp + \mathbf{r}_\perp$. For a topological charge $l = 1$ the interaction V is evidently a sum of ($\mathbf{r}_\alpha\, \mathbf{R}_\perp$) and ($\mathbf{r}_\alpha\, \mathbf{r}_\perp$). Application of the triangle-rule for the product of two dipoles, ($\mathbf{r}_\alpha\, \mathbf{r}_\perp$) say, tells us that the it can be represented by the sum of a scalar, dipole and a quadrupole $\Re^2_\mu(\mathbf{r})$. An expansion of $\Re^l_n(\mathbf{b})$ in products $\Re^a_\alpha(\mathbf{R}_\perp/w)$ & $\Re^c_\chi(\mathbf{r}_\perp/w)$ with $c \leq l$, where $a + \alpha$ & $c + \chi$ are even integers, facilitates the evaluation of matrix elements for $l \geq 2$.

Returning to the amplitude, we consider a typical term in F that is diagonal with respect to the topological charge. The product of the interesting matrix elements is,

$$F = \langle \lambda | \mathbf{r} \cdot \boldsymbol{\epsilon}' \{ \Re^l_n(\mathbf{b}) \}^* | \eta \rangle \langle \eta | \mathbf{r} \cdot \boldsymbol{\epsilon}\, \Re^l_n(\mathbf{b}) | \lambda' \rangle = \sum_{k',k} \sum_{K',K''} \sum_{K,Q} (2k'+1)(2k+1)\, \Upsilon^K_Q(k', k)$$

$$\times (-1)^{n+Q} \begin{pmatrix} 1 & l & k' \\ 0 & 0 & 0 \end{pmatrix} \begin{pmatrix} 1 & l & k \\ 0 & 0 & 0 \end{pmatrix} \{ \Pi^{K'} \otimes X^{K''} \}^K_Q [(2K'+1)(2K''+1)]^{1/2} \begin{Bmatrix} K' & K'' & K \\ l & 1 & k' \\ l & 1 & k \end{Bmatrix}, \tag{A6}$$

where $\Pi^{K'}_{Q'} = (ln\, l{-}n | K'\, Q')$ that is different from zero when $Q' = 0$. We assume that the intermediate state is spatially isotropic, to a good approximation, leaving it characterized solely by total angular momentum $J_c$ that resides in the atomic tensor $\Upsilon^K_Q(k', k)$. This simplification of the product of matrix elements is not necessary, however. A general result, with all quantum labels of the intermediate state, is given by Balcar and Lovesey together with steps in its reduction to (A6) [22]. The spherical tensor $\Upsilon^K_Q(k', k)$ is also a function of quantum labels in $|\lambda\rangle$ and $|\lambda'\rangle$ that belong to the ground-sate of an ion, whereas intermediate states $|\eta\rangle$ are virtual and do not obeys Hund's rules. Not shown explicitly in (A6) is a product of reduced matrix elements (RMEs) for spherical harmonics $[(l_v\|C(k')\|l_c)(l_c\|C(k)\|l_v)]$, where $l_v$ and $l_c$ are angular momenta for the valence and core states, respectively. An RME of this type is different from zero for $l_v + l_c + k$ even, say, so the aforementioned product is different from zero for $(k + k')$ even. The 3-j symbols in (A6) are different from zero for $(l + k')$ and ($l$

+ k) odd integers, which leads to the same condition on (k + k'). Variables in each row and each column of the 9-j symbol are subject to a triangular condition.

The Clebsch-Gordan coefficient $\Pi^{K'}{}_0 = (ln\ l{-}n | K'\ 0) = (-1)^{K'} (l{-}n\ ln | K'\ 0)$, i.e., $\Pi^{K'}{}_0$ is an odd function of $n$ for $K'$ odd and an even function of $n$ for $K'$ even. In an experiment this finding translates to a powerful selection rule on atomic information available from a difference ΔF of dichroic signals produced with opposite handedness in the photon beam. The selection rule becomes even more influential when it is combined with specific polarization in the primary beam, e.g., $K'' = 1$ for circular polarization.

The photon tensor for a twisted beam ($l = 1$) and circular polarization can be different from zero for zero projection ($Q = 0$), and we write it as $H^K{}_0(n, P_2)$. One finds,

$$H^K{}_0(+,+) = -\sqrt{(2K+1)} \begin{pmatrix} 2 & 2 & K \\ -2 & 2 & 0 \end{pmatrix} = (-1)^K H^K{}_0(-,-),$$

and,

$$H^K{}_0(+,-) = H^K{}_0(-,+) = -(1/6)\sqrt{(2K+1)} \begin{pmatrix} 2 & 2 & K \\ 0 & 0 & 0 \end{pmatrix},$$

is different from zero for $K$ even. Specific values of $H^K{}_0(n, P_2)$ appear in Table 4. The result $H^4{}_0(+,+) = H^4{}_0(+,-)$ accounts for the absence of a hexadecapole in the difference signal listed in Table 1.

For dichroism created with topological charge $l = 1$, application of the triangular condition shows that the rank $K' = 0, 1, 2$. Discussions in the main text concern quadrupole events and $k = k' = 2$ in (A6). Electronic multipoles then obey $0 \leq K \leq 4$, and in the application to dichroic signals $\Upsilon^K{}_Q(k', k)$ reduces to a multipole $\langle T^K{}_Q \rangle$ associated with the electronic ground sate, even though it depends on the total angular momentum of the core state $J_c$. The spherical tensor operator is Hermitian and $\langle T^K{}_Q \rangle^* = (-1)^Q \langle T^K{}_{-Q} \rangle$. The atomic multipole is completely specified by its RME (equation (73) in reference [4]), multiplied by $(l_v \| C(2) \| l_c)^2$ in an application [4]. The RME uses a standard unit-tensor that contains fractional parentage coefficients, and the unit-tensors have been listed for d and f atomic states [20]. A dependence on $J_c$ creates sum rules for integrated signals [2, 4].

| Process | Tensor | Prefactor | Projection (Q) | | | | |
|---|---|---|---|---|---|---|---|
| | | | 0 | ±1 | ±2 | ±3 | ±4 |
| Normal absorption | $X_Q^0$ | $-\dfrac{1}{\sqrt{3}}$ | 1 | | | | |
| E1-E1 | $X_Q^1$ | 0 | | | | | |
| Linear polarization ($x$) | $X_Q^2$ | $-\dfrac{1}{\sqrt{6}}$ | 1 | 0 | $-\sqrt{3/2}$ | | |
| Normal absorption | $H_Q^0$ | $\dfrac{1}{2\sqrt{5}}$ | 1 | | | | |
| E2-E2 | $H_Q^1$ | 0 | | | | | |
| Linear polarization ($x$) | $H_Q^2$ | $-\dfrac{1}{2\sqrt{14}}$ | 1 | 0 | $\sqrt{3/2}$ | | |
| | $H_Q^3$ | 0 | | | | | |
| | $H_Q^4$ | $-\dfrac{2}{\sqrt{70}}$ | 1 | 0 | $-\dfrac{\sqrt{10}}{4}$ | 0 | 0 |
| Normal absorption | $X_Q^0$ | $-\dfrac{1}{\sqrt{3}}$ | 1 | | | | |
| E1-E1 | $X_Q^1$ | $-\dfrac{1}{\sqrt{2}}$ | | ±1 | 0 | | |
| Circular polarization $P_2 \pm 1$ | $X_Q^2$ | $-\dfrac{1}{\sqrt{6}}$ | 1 | 0 | 0 | | |
| Normal absorption | $H_Q^0$ | $\dfrac{1}{2\sqrt{5}}$ | 1 | | | | |
| E2-E2 | $H_Q^1$ | $-\dfrac{1}{2\sqrt{10}}$ | | ±1 | 0 | | |
| Circular polarization $P_2 = \pm 1$ | $H_Q^2$ | $-\dfrac{1}{2\sqrt{14}}$ | 1 | 0 | 0 | | |
| | $H_Q^3$ | $\dfrac{1}{\sqrt{10}}$ | | ±1 | 0 | 0 | |
| | $H_Q^4$ | $-\dfrac{2}{\sqrt{70}}$ | 1 | 0 | 0 | 0 | 0 |

**Table 2**. Photon tensor components $\mathbf{X}^K$ (dipole transitions, equation (A2)) and $\mathbf{H}^K$ (quadrupole transitions, equations (1) or (4)) for normal absorption with linear polarization along the x-axis using $\boldsymbol{\epsilon} = \boldsymbol{\epsilon}' = (1, 0, 0)$. Polarization vectors $\boldsymbol{\epsilon} = (1, i, 0)/\sqrt{2}$ & $\boldsymbol{\epsilon}' = (1, -i, 0)/\sqrt{2}$ for right-handed circular polarization with Stokes parameter $P_2 = +1$. The photon wavevector is along the z-axis in Fig. 1a ($\hat{\boldsymbol{q}} = (0,0,1)$).

| Process | Tensor | Prefactor | Projection (Q) | | | | |
|---|---|---|---|---|---|---|---|
| | | | 0 | ±1 | ±2 | ±3 | ±4 |
| Borrmann Effect (E1-E1) | $X_Q^0$ | 0 | | | | | |
| Borrmann Effect (E2-E2) Linear polarization | $H_Q^0$ | $\dfrac{-1}{2\sqrt{5}}$ | 1 | | | | |
| | $H_Q^1$ | 0 | | | | | |
| | $H_Q^2$ | $\dfrac{1}{\sqrt{14}}$ | 1 | 0 | 0 | | |
| | $H_Q^3$ | 0 | | | | | |
| | $H_Q^4$ | $\dfrac{-1}{2\sqrt{70}}$ | 1 | 0 | 0 | 0 | $-\dfrac{\sqrt{70}}{2}$ |

**Table 3**. The photon tensor $H^K{}_Q$ is derived from either (1) or (4). Photon tensor components for the Borrmann case, with linear polarization along the x-axis using $\boldsymbol{\epsilon} = \boldsymbol{\epsilon}' = (1, 0, 0)$, $\hat{\mathbf{q}} = \boldsymbol{\kappa}_b = (0, 1, 0)$ and $\hat{\mathbf{q}}' = -\boldsymbol{\kappa}_b$.

| Process | Tensor | Prefactor | Projection (Q) | | | | |
|---|---|---|---|---|---|---|---|
| | | | 0 | ± | ±2 | ±3 | ±4 |
| OAM (E1-E1, \|n\|>0) | $X_Q^K$ | 0 | | | | | |
| OAM (E2-E2, $n = \pm 1$) | $H_Q^0$ | $\dfrac{-7}{12\sqrt{5}}$ | 1 | 0 | | | |
| E2-E2 | $H_Q^1$ | $\dfrac{1}{\sqrt{10}}$ | ±1 | 0 | | | |
| Linear polarization | $H_Q^2$ | $\dfrac{-5}{6\sqrt{14}}$ | 1 | 0 | $-\dfrac{\sqrt{6}}{5}$ | | |
| | $H_Q^3$ | $\dfrac{1}{2\sqrt{10}}$ | ±1 | 0 | $\mp\sqrt{5/6}$ | 0 | |
| | $H_Q^4$ | $\dfrac{-1}{\sqrt{70}}$ | 1 | 0 | $-\dfrac{\sqrt{10}}{4}$ | 0 | 0 |
| OAM circular polarization | $H_Q^0$ | $\dfrac{-1}{\sqrt{5}}$ | 1 | | | | |
| E2-E2 | $H_Q^1$ | $\dfrac{2}{\sqrt{10}}$ | ±1 | 0 | | | |
| $P_2 = \pm 1$  $n = \pm 1$ | $H_Q^2$ | $\dfrac{-2}{\sqrt{14}}$ | 1 | 0 | 0 | | |
| | $H_Q^3$ | $\dfrac{1}{\sqrt{10}}$ | ±1 | 0 | 0 | 0 | |
| | $H_Q^4$ | $\dfrac{-1}{\sqrt{70}}$ | 1 | 0 | 0 | 0 | 0 |
| OAM circular polarization | $H_Q^0$ | $\dfrac{-1}{6\sqrt{5}}$ | 1 | | | | |
| E2-E2 | $H_Q^1$ | 0 | | | | | |
| $P_2 = \pm 1$  $n = \mp 1$ | $H_Q^2$ | $\dfrac{1}{3\sqrt{14}}$ | 1 | 0 | 0 | | |
| | $H_Q^3$ | 0 | | | | | |
| | $H_Q^4$ | $\dfrac{-1}{\sqrt{70}}$ | 1 | 0 | 0 | 0 | 0 |

**Table 4**. The photon tensor $H^K_Q$ for the OAM (twisted beam) case, with linear polarization (top), and circular polarization parallel and antiparallel to the OAM (middle and bottom). The effective wave vectors for winding number $n = (\pm 1)$, are $\hat{\mathbf{q}} = \boldsymbol{\kappa}_t$ and $\hat{\mathbf{q}}' = -\boldsymbol{\kappa}_t^*$ with $\boldsymbol{\kappa}_t = (-i, 1, 0)/\sqrt{2}$ for $n = +1$, and $\boldsymbol{\kappa}_t = (i, 1, 0)/\sqrt{2}$ for $n = -1$. Circular polarization $P_2 = \pm 1$ with vectors $\boldsymbol{\epsilon} = (1, i, 0)/\sqrt{2}$ & $\boldsymbol{\epsilon}' = (1, -i, 0)/\sqrt{2}$ for right-handed circular polarization $P_2 = +1$.